\documentclass[epj]{svjour}
\usepackage{graphicx}
\usepackage{amsmath}
\usepackage{amssymb}
\usepackage{widetext}
\begin{document}
\title{{Hidden conformal symmetry for dyonic Kerr-Sen black hole and its gauged family}
\author{Muhammad Fitrah Alfian Rangga Sakti\thanks{\emph{e-mail:} fitrahalfian@gmail.com} 
}                     
%
%
\institute{High Energy Physics Theory Group, Department of Physics, Faculty of Science, Chulalongkorn University, Bangkok 10330, Thailand}
\date{Received: date / Revised version: date}
%
\abstract{
Motivated by advanced progress in the holographic theory between rotating black holes and CFT, we explore the conformal invariance on dyonic Kerr-Sen black hole and its gauged family. We consider a neutral massless scalar probe on the black holes' background within the low-frequency limit and exhibit that the solution space possesses $SL(2,R)\times SL(2,R)$ isometry. The periodic identification of the azimuthal angle corresponds to the spontaneous conformal symmetry breaking by temperatures $T_L,T_R$. Using the computation of the central charges on Ref.~\cite{SaktidyonicKerrSen} that we recalculate by considering the contributions of all associated fields, we successfully derive the Bekenstein-Hawking entropy from Cardy entropy formula. Furthermore, we also calculate the absorption cross-section from gravity side for generic non-extremal dyonic Kerr-Sen black hole and near-extremal gauged dyonic Kerr-Sen black hole. Our calculations show that those quantities are in a perfect match with the calculation from CFT. Therefore, our findings further support the duality between rotating black holes and CFTs.
}
\PACS{
      {04.40.Nr}{Einstein-Maxwell spacetimes, spacetimes with fluids, radiation or classical fields}   \and
      {04.60.-m}{Quantum gravity} \and
      {04.70.Dy}{Quantum aspects of black holes, evaporation, thermodynamics}
     } 
} 

\maketitle
\titlerunning{Hidden conformal symmetry}
\authorrunning{M.F.A.R. Sakti}

\section{Introduction}
\label{sec:intro}
It has been conjectured in the original Kerr/CFT correspondence that quantum gravity in the near-horizon region of extremal Kerr black hole is holographically dual to two-dimensional (2D)  conformal field theory (CFT) with a certain boundary condition \cite{GuicaPRD2009}. This conjecture has been explored extensively in the last decade for many black hole solutions in different theories, for example in low energy limit of heterotic string theory \cite{Ghezelbash2009}, in Einstein-Maxwell-Dilaton-Axion theory \cite{Li2010}, and in Rastall gravity \cite{SaktiAnnPhys2020,SaktiPhysDarkU2021}. So far, these calculations show that the near-horizon extremal geometry possesses $SL(2,R)\times U(1)$  isometry group where the $U(1)$ symmetry on the asymptotic infinity gets enhanced into Virasoro algebra. Furthermore, one can find another copy of Virasoro algebra from a different analysis on the boundary conditions as given in Refs.~\cite{MatsuoNucPhysB2010,CastroLarsenJHEP2009}. The isometry emerging exactly from the spacetime metric denotes the existence of warped AdS structure. The realization for the support of this Kerr/CFT conjecture is the precise agreement between Bekenstein-Hawking entropy and Cardy's growth of states in 2D CFT.

For the non-extremal rotating black holes, the near-horizon geometry has no global AdS structure. However, interestingly, some arbitrary symmetries are hidden in the solution space of the probe scalar field on the black hole's background. The first revelation of the hidden conformal symmetry was found in Ref.~\cite{Castro2010}. The conformal invariance is shown to appear as quadratic Casimir operators for the scalar probe possessing $SL(2,R)\times SL(2,R)$ isometry. This hidden conformal invariance has been explored also for extremal rotating black holes \cite{ChenLong2010,ChenWang2010} and near-extremal black holes \cite{ChenGhezelbash2011,SaktiNucPhysB2020}. Fascinatingly, one also can find the similar hidden conformal invariance for deformed scalar wave equation \cite{GhezelbashSiahaanCQG2013,GhezelbashSiahaanGRG2014,Saktideformed2019}. The extension for this exploration of the hidden conformal symmetry for higher spin fields can be found in Ref.~\cite{LoweSkanataPRD2014}. Moreover, this conformal invariance appears also on the horizonless compact object resulting in similar CFT prescription \cite{DeyAfshordiPRD2020,SaktiPhysDarkU2022}. For the review of the Kerr/CFT correspondence in extremal, near-extremal, and non-extremal cases, one can see in Ref. \cite{Compere2017}.

In the study of hidden conformal symmetry, the computation of the central charges is carried out by numerologically observing the Bekenstein-Hawking entropy in order to find the agreement with Cardy formula. Nevertheless, recently, another significant advance for the computation of the central charges for the non-extremal background has been discovered. In Ref.~\cite{HacoJHEP2018}, they have exhibited explicitly a full set of $Vir_{L}\times Vir_{R}$ symmetries for the non-extremal Kerr black holes. The Wald-Zoupas boundary counter term is used to remove certain obstructions that leads to the discovery of non-vanishing left- and right-moving central charges. This calculation eventually has proven the numerological observation done in the original work of hidden conformal symmetry \cite{Castro2010}. In that work, they have generalized the work on Schwarzschild black holes that implements a different class of symmetry generators \cite{AveninPRD2020}. The success of these works is then generalized for generic Killing horizons \cite{ChenTorresPRL2020}. The near-horizon expansion in conformal coordinates for this generic Killing horizons shows that, on the bifurcation surface, the spacetime metric possesses a locally $AdS_3$ structure. This has suggested some more studies on the dual CFT description for arbitrary Killing horizons.

Motivated by the previous advanced study on the conformal invariance of the rotating black holes, in this paper we investigate hidden conformal invariance for dyonic rotating black holes in Einstein-Maxwell-Dilaton-Axion (EMDA) theory, namely dyonic Kerr-Sen and its gauged family. This study is also the extension of the results in Ref. \cite{GhezelbashSiahaanCQG2013} for Kerr-Sen black holes. We want to extend our previous study where we have carried out the conformal symmetry directly on the spacetime metric for near-horizon extremal dyonic rotating black hole in Einstein-Maxwell-Dilaton-Axion (EMDA) theory, its gauged, and ultraspinning families \cite{SaktidyonicKerrSen}. The presence of the dyonic/magnetic charge, as well as the gauge coupling constant make a rich structure for the horizons of the black holes, which in turn, needs careful analysis to reveal the conformal structure for the wave equation of a neutral massless scalar probe in the background of the black holes. We also calculate the scattering cross-section of the scalar field in the near region of the non-extremal dyonic Kerr-Sen black hole, as well as the near-extremal gauged Kerr-Sen black hole. We also discuss the real-time correlators for both backgrounds of the black hole solutions.

The organization of this paper is given as follows. After giving the introduction in Sec. \ref{sec:intro}, we consider two main black hole solutions from EMDA theory that will be studied in Sec. \ref{sec:bhspacetime}. In Sect. \ref{sec:central}, we give the calculation of the central charge by considering the contributions from all fields. In Sec. \ref{sec:hidden}, we consider a neutral massless scalar probe on the background of dyonic Kerr-Sen and gauged dyonic Kerr-Sen black holes. We study the hidden conformal symmetries of the solution space of the scalar probes for both black holes. The entropy is computed using Cardy prescription of 2D CFT. The scattering issues of the scalar probes are then investigated in Sec. \ref{sec:scat} to obtain the absorption cross-sections. We discuss the absorption cross-section for generic non-extremal dyonic Kerr-Sen black hole and near-extremal gauged dyonic Kerr-Sen black hole from gravity and CFT computations. We also provide the analysis of real-time correlators that correspond with the retarded Green's function. In the end, we summarize the whole paper in Sec. \ref{sec:conc}.

\section{Black Holes Spacetimes and Their Properties}
\label{sec:bhspacetime}
\subsection{Dyonic Kerr-Sen black hole}
The dyonic Kerr-Sen spacetime is the solution of the Einstein-Maxwell-Dilaton-Axion (EMDA) theory,
\begin{eqnarray}
\mathcal{L}&=&\sqrt{-g}\left[R - \frac{1}{2}(\partial\phi)^2-\frac{1}{2}e^{2\phi}(\partial\chi)^2-e^{-\phi}F^2\right] \nonumber\\
&+&\frac{\chi}{2}\epsilon^{\mu\nu\rho\lambda}F_{\mu\nu}F_{\rho_\lambda}, \label{eq:Lagrangian}\
\end{eqnarray}
where $R$ is the Ricci scalar, $F=dA$ is the Maxwell tensor, and the last term is the topological term. $\phi$ is the dilaton field while $\chi$ is the (pseudoscalar) axion field. This Lagrangian can be written to another form as given in low energy limit of heterotic strng theory (see Appendix \ref{app:1}). The spacetime metric is given in Ref.~\cite{WuWuWuYuPRD2021}. In this paper, we follow the spacetime metric form in shifted radial coordinate which is exactly given in Ref.~\cite{SaktidyonicKerrSen}. The black hole spacetime metric is given by
\begin{equation}
	ds^2 = - \frac{\Delta}{\varrho^2 }X^2+  \frac{\varrho ^2}{\Delta}dr^2 + \varrho ^2 d\theta^2 + \frac{\sin^2\theta}{\varrho ^2}Y^2,\label{eq:dKSmetric} \
\end{equation}
respectively, where
\begin{eqnarray}
X &=&dt - a \sin^2\theta d\phi, \nonumber\\
Y &=& adt- (r^2-d^2-k^2+a^2 ) d\phi, \nonumber\\
\varrho^2 &=& r^2-d^2 -k^2 + a^2\cos^2\theta , \nonumber\\
\Delta &=& r^2 -2m r -d^2 -k^2+a^2 +p^2 +q^2. \label{eq:metricfunctiondKS}\
\end{eqnarray}
The electromagnetic field, its dual, dilaton, and axion fields related to metric (\ref{eq:dKSmetric}) are given by
\begin{equation}
\textbf{A} = \frac{q(r+d-p^2/m)}{\varrho^2}X-\frac{p\cos\theta}{\varrho^2}Y, \label{eq:AdKS}\
\end{equation}
\begin{equation}
\textbf{B} = \frac{p(r+d-p^2/m)}{\varrho^2}X+\frac{q\cos\theta}{\varrho^2}Y, \label{eq:BdKS}\
\end{equation}
\begin{equation}
e^{\phi} = \frac{(r+d)^2+(k+a\cos\theta)^2}{\varrho^2}, \label{eq:dilatondKS}\
\end{equation}
\begin{equation}
\chi = 2\frac{kr-da\cos\theta}{(r+d)^2+(k+a\cos\theta)^2}.\ \label{eq:axiondKS}
\end{equation}
The dual gauge potential can be obtained from $-dB=e^{-\phi}\star F+\chi F$. The parameters $m, a, q, p, d, k$ are mass, spin, electric charge, magnetic (dyonic) charge, dilaton charge, and axion charge of the black hole, respectively. For non-dyonic solution, the solution has been carried out in Ref.~\cite{WuWuYuPRD2020}. The dilaton and axion charges explicitly depend on the electromagnetic charges with the following relations
\begin{equation}
	d= \frac{p^2 -q^2}{2m}, ~~~~~ k =\frac{pq}{m}.\
\end{equation}

The thermodynamic relation of dyonic Kerr-Sen black hole has been investigated in Ref.~\cite{WuWuWuYuPRD2021}. The Hawking temperature, Bekenstein-Hawking entropy, angular velocity, eletric potential, and magnetic potential of the black hole solution are given by
\begin{equation}
	T_H = \frac{r_+ -m}{2\pi (r_+^2 -d^2-k^2 +a^2)}, \label{eq:THKS}
\end{equation}
\begin{equation}
	S_{BH}=\pi(r_+^2 -d^2 -k^2 +a^2), \label{eq:SBHKS}
\end{equation}
\begin{equation}
\Omega_H = \frac{a}{r_+^2 -d^2 -k^2 +a^2},
\end{equation}
\begin{equation}
\Phi_H = \frac{q(r_+ +d -p^2/m)}{r_+^2 -d^2 -k^2 +a^2}, \label{eq:potphiKS}
\end{equation}
\begin{equation}
\Psi_H = \frac{p(r_+ +d -p^2/m)}{r_+^2 -d^2 -k^2 +a^2}. \label{eq:potpsiKS}
\end{equation}
The position of the inner and outer horizons are as given by
\begin{equation}
	r_\pm = m \pm \sqrt{ m^2 + d^2 + k^2 -a^2 - p^2 - q^2}.
\end{equation}
Above thermodynamic quantities are studied in the macroscopic point of view. The microscopic description of the thermodynamic quantities can also be studied extensively using different novel methods such as Kerr/CFT correspondence \cite{GuicaPRD2009}. We have applied the Kerr/CFT method to this black hole in Ref.~\cite{SaktidyonicKerrSen} to calculate the entropy using Cardy's growth of states for 2D CFT to extremal dyonic Kerr-Sen black hole. Note that in extremal case, the position of the event horizon is located on $r_+ = r_- = m$. To reproduce Cardy entropy, we have applied the asymptotic symmetry group to calculate the central charge while the temperature has been computed using generalized Frolov-Thorne vacuum. It has been found that, enticingly the central charge possesses two branches given by \cite{SaktidyonicKerrSen}
\begin{equation}
	c_L = 12am_\pm, \label{eq:cLgdyonicKS}\
\end{equation}
where the mass is given by
\begin{equation}
	m_\pm^{2} = \frac{1}{2}(a^{2}+p^{2}+q^{2})\left[ 1\pm \sqrt{1-\left( \frac{p^{2}+q^{2}}{a^{2}+p^{2}+q^{2}}\right)^{2}} \right]. \label{eq:mext}
\end{equation}
The mass will reduce into one only when the spin reaches zero implying that the black hole is not rotating anymore. This also results in vanishing entropy like for the case of GMGHS black hole \cite{HerdeiroPLB2021}. The left-moving temperature for the input to Cardy formula is the given by \cite{SaktidyonicKerrSen}
\begin{equation}
	T_L =  \frac{m_\pm^2-d^2-k^2+a^2}{4\pi a m_\pm}. \label{eq:TLKS}\
\end{equation}
The final result is the Cardy entropy
\begin{equation}
	S_{CFT}=\pi(m_\pm^2 -d^2 -k^2 +a^2), \label{eq:SCFTKSAdSshift}
\end{equation}
which is equal to Bekenstein-Hawking entropy. When we assume $q=0$, we recover the results from extremal Kerr-Sen black hole. These results also recover the entropy of extremal Kerr black hole when all electromagnetic charges vanish.

\subsection{Gauged Dyonic Kerr-Sen black hole}
The gauged version of the dyonic Kerr-Sen black hole or the Kerr-Sen-AdS black is given by \cite{SaktidyonicKerrSen} which is the solution of the following action
	\begin{eqnarray}
	\mathcal{L}_{gauged} =\mathcal{L} + \sqrt{-g}\frac{4+e^{-\phi}+e^{\phi}(1+\chi^2)}{l^2}.\label{eq:Lagrangian1}\
\end{eqnarray}
The spacetime metric is given by
\begin{equation}
	ds^2 = - \frac{\Delta}{\varrho^2 } X^2+  \frac{\varrho ^2}{\Delta}dr^2 + \frac{\varrho ^2}{\Delta_\theta} d\theta^2 + \frac{\Delta_\theta \sin^2\theta}{\varrho ^2}Y^2,\label{eq:dKSAdSmetricshift} \
\end{equation}
where
\begin{eqnarray}
	X &=& dt - a \sin^2\theta \frac{d\phi}{\Xi}, \nonumber\\
	 Y &=& adt- (r^2-d^2-k^2+a^2 ) \frac{d\phi}{\Xi}, \nonumber\\
	\Delta &=& (r^2-d^2-k^2+a^2)\left(1+\frac{r^2-d^2 -k^2}{l^2} \right), \nonumber\\
	& & -2mr+p^2+q^2 \nonumber\\
	\Delta_\theta &=& 1-\frac{a^2}{l^2}\cos^2\theta, ~~\Xi =1-\frac{a^2}{l^2}, \nonumber\\
	\varrho^2 &=& r^2-d^2 -k^2 + a^2\cos^2\theta.\label{eq:metricfunctiondKSAdS}\
\end{eqnarray}
The electromagnetic field, its dual, dilaton, and axion fields are given by Eqs. (\ref{eq:AdKS})-(\ref{eq:axiondKS}) but with metric functions (\ref{eq:metricfunctiondKSAdS}). From this gauged version of dyonic Kerr-Sen black hole solution, several solutions can be obtained by taking some limits. The Kerr-Sen-AdS black hole can be obtained when one turns off $p=0$ that also will cause $k=0$. When one consider equal charges $q=p$, dilaton charge will vanish. When we assume that $q=p=0$, we can find the Kerr-AdS black hole solution.

When the gauge coupling constant $l$ arises as the vacuum expectation value, this parameter can include in the first law of thermodynamics for black holes \cite{CveticGibbonsPRD2011}. The gauge coupling constant can be considered as the source of the pressure on the black holes, giving rise to another thermodynamic quantity. All of the thermodynamic quantities of the gauged dyonic Kerr-Sen black hole is given by
\begin{equation}
M= \frac{m}{\Xi}, ~~~ J = \frac{ma}{\Xi}, ~~~ Q= \frac{q}{\Xi}, ~~~ P= \frac{p}{\Xi}, \label{eq:MJQPKSAdS}
\end{equation}
\begin{equation}
T_H = \frac{r_+(2r_+^2-2d^2 -2k^2+a^2 +l^2)-ml^2}{2\pi (r_+^2 -d^2-k^2 +a^2)l^2}, \label{eq:THKSAdS}
\end{equation}
\begin{equation}
S_{BH}=\frac{\pi}{\Xi}(r_+^2 -d^2 -k^2 +a^2), \label{eq:SBHKSAdS}\
\end{equation}
\begin{equation}
\Omega_H = \frac{a\Xi}{r_+^2 -d^2 -k^2 +a^2}, \label{eq:OmegaKSAdS}\
\end{equation}
\begin{equation}
\Phi_H = \frac{q(r_+ +d -p^2/m)}{r_+^2 -d^2 -k^2 +a^2}, \label{eq:PhiKSAdS}
\end{equation}
\begin{equation}
\Psi_H = \frac{p(r_+ +d -p^2/m)}{r_+^2 -d^2 -k^2 +a^2}, \label{eq:PsiKSAdS}\
\end{equation}
\begin{equation}
V= \frac{4}{3}r_+ S, ~~~\mathcal{P}=\frac{3}{8\pi l^2}. \label{eq:VdKSAdS}
\end{equation}
Those quantities satisfy the following thermodynamic relation \begin{eqnarray}
	dM= T_H dS+\Omega_H dJ +\Phi_H dQ +\Psi_H dP + Vd\mathcal{P}. \label{eq:thermoKSAdS}
\end{eqnarray}
The event horizon $r_+$ for this gauged case is different with non-gauged one. In this gauged case, the cosmological constant is considered as a thermodynamic variable expressed in terms of $\mathcal{P}$. This leads to the conjecture that this asymptotically AdS black hole solution may violate the isoperimetric inequality \cite{CveticGibbonsPRD2011,DolanKastorPRD2013}. For this gauged dyonic Kerr-Sen black hole, it depends on the value of $d,k$ \cite{WuWuWuYuPRD2021}. When it violates the inequality, the entropy of this black hole is maximized. Furthermore, proposing cosmological constant as another thermodynamic variable is also based on the study where mass of the black hole should be considered as enthalpy of the spacetime. The consequence of this is we must add the cosmological constant term analogous to pressure in Smarr formula \cite{CveticGibbonsPRD2011,DolanKastorPRD2013}.

The Kerr/CFT calculation for this extremal gauged family has also been applied in Ref.~\cite{SaktidyonicKerrSen}. The resulting central charge for this black hole solution is given by
\begin{equation}
	c_L = \frac{12ar_+}{\upsilon},\label{eq:cLgeneral}
\end{equation}
where
\begin{equation}
\upsilon = 1 + \frac{6r_+^2-2d^2-2k^2+a^2}{l^2}. \label{eq:upsilonKSAdS}\
\end{equation}
All of the horizons coincides into one $r_+$ for extremal black holes. Fascinatingly, it has been noted in Ref.~\cite{SaktidyonicKerrSen} that for this gauged black hole, there possibly more than two branches of the central charge in terms of mass since $d,k\sim 1/m$. The left-moving temperature is then given by
\begin{equation}
	T_L =  \frac{\upsilon(r_+^2-d^2-k^2+a^2)}{4\pi a r_+ \Xi}. \label{eq:TLgeneral}\
\end{equation}
As the final result, using Cardy entropy formula it has been found that
\begin{equation}
	S_{CFT}=\frac{\pi}{\Xi}(r_+^2 -d^2 -k^2 +a^2). \label{eq:SCFTKSAdSshift}
\end{equation}
When we assume $q=0$, we recover the results from extremal gauged Kerr-Sen black hole. These results also recovers the entropy of extremal Kerr-AdS black hole when all electromagnetic charges vanish.

\section{Central Charge}
\label{sec:central}
In the previous section, we have shown the central charges obtained in Ref. \cite{SaktidyonicKerrSen} fro dyonic Kerr-Sen black hole and its gauged family. However, in Ref. \cite{SaktidyonicKerrSen}, only the contribution from the graviton has been considered explicitly. The objective of this section is to re-calculate the central charge computed in Ref. \cite{SaktidyonicKerrSen} for dyonic Kerr-Sen metric (\ref{eq:dKSmetric}) and its gauged family (\ref{eq:dKSAdSmetricshift}). In this section, we will perform the computation of the conserved quantities that lead to the central charges associated to the diffeomorphisms of all fields in EMDA theory. 

 We will compute the diffeomorphisms and the conserved charges related to near-horizon extremal gauged dyonic Kerr-Sen black hole given in Eq. (4.14) in Ref. \cite{SaktidyonicKerrSen},
\begin{eqnarray}
	ds^2 &=& \Gamma(\theta)\left(-r^2 dt^2 + \frac{dr^2}{r^2} + \alpha(\theta) d\theta ^2 \right) \nonumber\\
	&& +\gamma(\theta) \left(d\phi +e r dt\right)^2, \label{eq:extremalmetricnewdKSAdS}\
\end{eqnarray}
where the metric functions are given by
\begin{eqnarray}
	&&\Gamma(\theta)= \frac{\varrho_+ ^2}{\upsilon} , ~~~\alpha(\theta) = \frac{\upsilon}{\Delta_\theta}, ~~~\gamma(\theta)=\frac{r_0^4 \Delta_\theta \sin^2\theta}{\varrho_+^2 \Xi^2}, \nonumber\\
	&&\varrho_+^2 = r_+^2 -d^2 -k^2 + a^2\cos^2\theta, ~~~
	e = \frac{2ar_+\Xi}{r_0^2 \upsilon}. \label{eq:constant_e}
\end{eqnarray}
The near-horizon gauge field is given by \cite{SaktidyonicKerrSen},
\begin{eqnarray}
	\textbf{A} &=& f_a(\theta) \left(d\phi +er dt \right)\nonumber\\
	&+& \frac{q\left[(r_+ +d)^2+k^2-a^2-2p^2r_+/m \right]}{\hat{e}(r_+^2-d^2-k^2+a^2)}d\phi,\\
	f_a(\theta) &=& \frac{q\left[\frac{2r_+p^2}{m}-(r_+ + d)^2-k^2+a^2\cos^2\theta\right] }{2ar_+\Xi \varrho_+^2}r_0^2,\nonumber\\
	&+&\frac{2apr_+\cos\theta}{2ar_+\Xi \varrho_+^2}r_0^2. \label{eq:nearhorizonelectromagneticpot1}
\end{eqnarray}
For the ungauged case, we just need to set $1/l^2=0$.

Instead of using the calculation provided in Ref. \cite{Ghezelbash2009}, we will perform the similar computation given in Ref. \cite{CompereMurataJHEP2009,NoorbakhshPRD2017} where they manage to compute the central charges of the gauged supergravity black holes. For this calculation we cannot write all related fields into the antisymmetric tensor field $\mathcal{B}$ or antisymmetric tensor $H$ as given in \cite{Ghezelbash2009}. The last term in (\ref{eq:Lagrangian}) as the topological term will give another contribution. Moreover, it has been pointed out also in \cite{BarnichBrandtNPB2002,BarnichCompereJMP2008} that cosmological constant term does not contain the derivatives of metric deviation $h_{\mu\nu}$, so it will not contribute on the superpotential of the central term. 

Now we recall the asymptotic symmetry group (ASG) of a spacetime which is the group of allowed symmetries which obey the imposed boundary conditions. To calculate the charges associated with ASG of gauged dyonic Kerr-Sen solution, we consider the asymptotic symmetries of the action (\ref{eq:Lagrangian1}) that includes the diffeomorphisms $\xi$ such that \cite{CompereMurataJHEP2009,NoorbakhshPRD2017}
\begin{eqnarray}
	&&\delta_{\xi}\phi=\partial\phi = \xi_\mu \nabla^\mu \phi,\\
	&&\delta_{\xi} A_\mu=a_\mu = \xi^\nu F_{\mu\nu}+\nabla_\mu(A_\nu\xi^\nu),\\
	&&\delta_{\xi} g_{\mu\nu}=h_{\mu\nu}=\nabla_\mu \xi_\nu + \nabla_\nu \xi_\mu,\
	\
\end{eqnarray}
as well as the gauge transformation related to the gauge field, graviton, and dilaton field, respectively,
\begin{eqnarray}
	\delta_\Lambda A_\mu = \partial_\mu \Lambda,~~\delta_\Lambda g_{\mu\nu}=0, ~~\delta_\Lambda\phi=0.
\end{eqnarray}
In order to compute the non-trivial diffeomorphisms, we must impose some arbitrary boundary conditions. In this case, we follow the boundary conditions from Ref. \cite{CompereMurataJHEP2009} for all fields for the near-horizon extremal metric (\ref{eq:extremalmetricnewdKSAdS}).

\begin{eqnarray}
	h_{\mu \nu} \sim \left(\begin{array}{cccc}
		\mathcal{O}(r^2) & \mathcal{O}\left(\frac{1}{r^2}\right) &  \mathcal{O}\left(\frac{1}{r}\right) &  \mathcal{O}(1) \\
		&  \mathcal{O}\left(\frac{1}{r^3}\right) &  \mathcal{O}\left(\frac{1}{r^2}\right) &  \mathcal{O}\left(\frac{1}{r}\right) \\
		&  & \mathcal{O}\left(\frac{1}{r}\right) &  \mathcal{O}\left(\frac{1}{r}\right)\\
		&  &  &  \mathcal{O}(1)\
	\end{array} \right),\label{eq:gdeviation}
\end{eqnarray}
\begin{equation}
a_\mu \sim \left(\mathcal{O}(r),\mathcal{O}\left(1/r^2\right),\mathcal{O}(1),\mathcal{O}(1/r)\right), \label{eq:adeviation}\
\end{equation}
\begin{equation}\partial\phi \sim \mathcal{O}(1), \label{eq:phideviation}\
\end{equation}
in the basis $(t,r,\theta,\phi)$. The boundary conditions on metric deviation is already implemented in most of the Kerr/CFT correspondence's papers (see \cite{Hartman2009,Ghezelbash2009,Lu2009,Astorino2015,Astorino2016,Sinamuli2016,Sakti2018,SaktiEPJPlus2019} for example). The most general diffemorphisms that preserve the boundary conditions (\ref{eq:gdeviation})-(\ref{eq:phideviation}) are given by
\begin{eqnarray}
	\xi_\epsilon = \epsilon_n(\phi)\partial_\phi - r\epsilon_n'(\phi)\partial_r, ~~ \bar{\xi}=\partial_t ,\label{eq:killingASG}
\end{eqnarray}
where $\epsilon_n(\phi)=-e^{-in\phi}$. The diffeomorphisms commute to each other. Above diffeomorphisms satisfy
\begin{eqnarray}
	i[\xi_m, \xi_n]_{LB} = (m-n)\xi_{m+n}.
\end{eqnarray}
Nonetheless, the boundary conditions (\ref{eq:adeviation}) are not satisfied under $\xi_\epsilon$ since we find $\delta_\epsilon A=f_a\epsilon' (d\phi-krdt)$ or $\delta_\epsilon A_\phi$ is of the order of 1 at infinity. So, we must add a compesating $U(1)$ gauge transformation to restore $\delta_\epsilon A_\phi=\mathcal{O}(1/r)$ as given by $\Lambda =-f_a(\theta)\epsilon(\phi)$. Now the asymptotic symmetries consist of $(\xi_n,\Lambda_n)$ satisfying
\begin{eqnarray}
	[\Lambda_n,\Lambda_m]&=&\xi_n^\mu\partial_\mu \Lambda_m-\xi_m^\mu\partial_\mu\Lambda_n,\\
	i[(\xi_n,\Lambda_n),(\xi_m,\Lambda_m)]&=&(n-m)(\xi_{n+m},\Lambda_{n+m}).\label{eq:Wittalgebrafull}\
\end{eqnarray}
The associated conserved charge is then
\begin{eqnarray}
	Q_{\xi,\Lambda} = \frac{1}{8\pi}\int_{\partial\Sigma} k^{tot}_{\xi,\Lambda},
\end{eqnarray}
where the superpotential $k^{tot}_{\xi,\Lambda}=k^{g}_{\xi}+k^F_{\xi,\Lambda}+k^{top}_{\xi,\Lambda}+k^\phi_\xi$ representing the contributions from graviton, electromagnetic field, topological term, and the dilaton field. The given integral is over the boundary of a spatial slice. The contributions of corresponding fields on the central charge are given explicitly by
\begin{eqnarray}
	k^{g}_{\xi}&=& \frac{1}{8\pi} \bigg\{ \xi ^{\nu} \nabla^{\mu} h - \xi ^{\nu} \nabla_{\sigma} h^{\mu \sigma} + \frac{h}{2} \nabla^{\nu}\xi^{\mu}\nonumber\\
	&-&  h^{\rho\nu} \nabla_{\rho}\xi^{\mu} + \xi_{\sigma}\nabla^{\nu}h^{\mu \sigma}  \nonumber\\
	&+& \left. \frac{h^{\sigma \nu}}{2}\left(\nabla^\mu \xi_\sigma + \nabla_\sigma \xi^\mu \right) \right\} dx^\mu \wedge dx^\nu, \label{eq:kgrav} \\
	k_{\xi,\Lambda}^{F}&=&\frac{1}{16\pi}\bigg\{\bigg(-k_\chi F^{\mu\nu}\delta\phi+2k_\chi h^{\mu\lambda}F_\lambda^\nu  \nonumber\\
	&-&\left.  k_{\chi} \delta F^{\mu\nu} -\frac{1}{2}h k_{\chi} F^{\mu\nu} \right)(A_\rho\xi^\rho+\Lambda) \nonumber\\
	&-&k_\chi F^{\mu\nu}a_\rho \xi^\rho -2\xi^\mu k_\chi F^{\nu\lambda} a_\lambda \nonumber\\
	&-&k_\chi a^\mu g^{\nu\sigma}(\mathcal{L}_\xi A_\sigma +\partial_\sigma \Lambda) \bigg\} dx^\mu \wedge dx^\nu,\label{eq:kF}\\
	k^{top}_{\xi\Lambda} &=& \frac{1}{8\pi}\bigg\{ \epsilon^{\mu\nu\lambda\sigma}(h_{\chi,A}F_{\lambda\sigma}\delta\phi +h_\chi \partial F_{\lambda\sigma})(A_\rho\xi^\rho+\Lambda) \nonumber\\
	&+&\epsilon^{\mu\nu\lambda\sigma}h_\chi F_{\lambda\sigma} a_\rho \xi^\rho - 2\xi^\nu h_\chi \epsilon^{\mu\lambda\rho\sigma}F_{\rho\sigma}a_\lambda \nonumber\\
	&-&2h_\chi\epsilon^{\mu\nu\rho\sigma}a_\rho (\mathcal{L}_\xi A_\sigma +\partial_\sigma \Lambda)\bigg\} dx^\mu \wedge dx^\nu, \label{eq:ktop} \\
	k^\phi_\xi &=&\frac{1}{8\pi}\xi^\nu \nabla^\mu \phi \delta\phi ~ dx^\mu \wedge dx^\nu,\label{eq:kphi} \
\end{eqnarray}
where $k_\chi= 4e^{-\phi}, h_\chi=\chi/2, \delta F_{\mu\nu}=\partial_\mu a_\nu-\partial_\nu a_\mu$.
We should note that the last two terms in Eqs. (\ref{eq:kgrav})-(\ref{eq:ktop}) vanish for an exact Killing vector and an exact symmetry, respectively. The charge $ Q_{\xi} $ generates symmetry through the Dirac brackets. The ASG possesses algebra which is given by the Dirac bracket algebra of the following charges \cite{BarnichBrandtNPB2002}
\begin{eqnarray}
	\{ Q_{\xi,\Lambda},Q_{\bar{\xi},\bar{\Lambda}}\}_{DB} &=& \frac{1}{8\pi}\int_{\partial\Sigma} \left(k^{g}_{\xi}\left[\mathcal{L}_{\bar{\xi}}g;g \right]\right.\nonumber\\
	&+&k^\phi_\xi[\mathcal{L}_{\bar{\xi}}g,\mathcal{L}_{\bar{\xi}}\phi;g,\phi]\nonumber\\
	&+&k^A_{\xi,\Lambda}\left[\mathcal{L}_{\bar{\xi}}g,\mathcal{L}_{\bar{\xi}}A+d\Lambda,\mathcal{L}_{\bar{\xi}}\phi;g,A,\phi\right]\nonumber\\
	&+&\left.k^{top}_{\xi,\Lambda}\left[\mathcal{L}_{\bar{\xi}}g,\mathcal{L}_{\bar{\xi}}A+d\Lambda,\mathcal{L}_{\bar{\xi}}\phi;g,A,\phi\right]\right)\nonumber\\
	&=& Q_{[(\xi,\Lambda),(\bar{\xi},\bar{\Lambda})]} + \left(k^{g}_{\xi}\left[\mathcal{L}_{\bar{\xi}}\bar{g};\bar{g} \right]\right.\nonumber\\
	&+&k^A_{\xi,\Lambda}\left[\mathcal{L}_{\bar{\xi}}\bar{g},\mathcal{L}_{\bar{\xi}}\bar{A}+d\Lambda,\mathcal{L}_{\bar{\xi}}\bar{\phi};\bar{g},\bar{A},\bar{\phi}\right]\nonumber\\
	&+&k^{top}_{\xi,\Lambda}\left[\mathcal{L}_{\bar{\xi}}\bar{g},\mathcal{L}_{\bar{\xi}}\bar{A}+d\Lambda,\mathcal{L}_{\bar{\xi}}\bar{\phi};\bar{g},\bar{A},\bar{\phi}\right]\nonumber\\
	&+&\left.k^\phi_\xi[\mathcal{L}_{\bar{\xi}}\bar{g},\mathcal{L}_{\bar{\xi}}\bar{\phi};\bar{g},\bar{\phi}]\right) \label{eq:charges},\
\end{eqnarray}
where $\bar{g},\bar{A},\bar{\phi}$ are the background solutions for each field. Because of $\delta_\xi \phi +\delta_\Lambda \phi=0$, the contribution from the dilaton field vanishes. Hence, $k^\phi_{\xi}=0$. Moreover, one can also obtain that $k^{F}=k^{top}=0$ by inserting $\xi,\Lambda$. On the other hand, $ c_A =c_{top}=c_\phi=0$. So, the remaining non-zero contribution to the central charge is from the gravity. One can find the algebra of the charge $Q_n$ associated to ASG generators ($\xi_n,\Lambda_n$) above Dirac bracket as
\begin{equation}
	i\{Q_m,Q_n\}_{DB}=(m-n)Q_{m+n}+\frac{c_L}{12}(m^3-xm)\delta_{m+n,0}.\
\end{equation}
The constant $x$ is just a constant that can be absorbed into $Q_0$. So, we obtain that
\begin{eqnarray}
	c_L=c_g=\frac{3e}{\Xi}\int^\pi_0d\theta\sqrt{\Gamma(\theta)\alpha(\theta)\gamma(\theta)}=\frac{12ar_+}{\upsilon}.\label{eq:cLtot}
\end{eqnarray}
This result is exactly similar with the result given in Ref.~\cite{SaktidyonicKerrSen} as shown in Eq. (\ref{eq:cLgeneral}). When the cosmological constant vanishes, we obtain that $\upsilon=1$, so we recover the central charge in Eq. (\ref{eq:cLgdyonicKS}).

\section{Hidden Conformal Symmetry and CFT Entropy }\label{sec:hidden}
\subsection{Dyonic Kerr-Sen black hole}
In this section, we are going to demonstrate that there is a hidden conformal symmetry on the dyonic Kerr-Sen black hole likewise its non-dyonic family \cite{GhezelbashSiahaanCQG2013}. Firstly, we need to find the radial equation of the scalar probe where the conformal symmetry is hidden. The massless scalar field equation for the scalar probe, is given by
\begin{equation}
	\nabla_{\alpha} \nabla^{\alpha}\Phi = 0\label{KG1}.
\end{equation}
We notice that the dyonic Kerr-Sen black hole (\ref{eq:dKSmetric}) have two translational Killing vectors i.e., $\partial_t$ and $\partial_\phi$. So, we can separate the coordinates in the scalar field as 
\begin{equation}
	\Phi(t, r, \theta, \phi) = \mathrm{e}^{- i \omega t + i m \phi} R(r) S(\theta)\label{phi-expand1}.
\end{equation}
It is worth to point out that we can also assume the charged scalar field in the wave equation. However, as noted in \cite{GhezelbashSiahaanCQG2013}, the Kerr-Sen black hole does not have charge picture. It means that when the scalar field is electromagnetically charged, the solution space does not shows well-defined hidden conformal symmetry. Nevertheless, this case is different for dyonic Kerr-Newman which possesses hidden conformal symmetry both in eletric and magnetic pictures \cite{ChenHuangPRD2010}. Plugging Eq. (\ref{phi-expand1}) into Eq. (\ref{KG1}), leads to two differential equations i.e., the angular $S(\theta)$ and radial $R(r)$ wave functions,
\begin{equation}
	\left[\frac{1}{\sin\theta} \partial_\theta (\sin\theta \partial_\theta )-  \frac{m^2}{\sin^2\theta} - a^2\omega^2 \sin^2\theta \right] S(\theta) =- K_h S(\theta)  \label{angular1}, \
\end{equation}
\begin{eqnarray} 
&&	\bigg[\partial_r (\Delta \partial_r) + \frac{\left[ (r^2 -d^2 -k^2 +a^2) \omega - am \right]^2}{\Delta} + 2 a m \omega \bigg] R(r) \label{radial1}, \nonumber\\
&&	= K_h R(r).\label{eq:radial1}\
\end{eqnarray}
where the separation constant $K_{h}$ is the eigenvalues on a sphere. 

The hidden conformal symmetries can be seen when we look at the near region ($r\ll 1/\omega$) of the scalar wave equation and also by neglecting some terms. Some terms can be neglected when we consider low frequency. In order to get so, we consider the low-frequency limit for the scalar field $ \omega M \ll 1 $. Consequently, we also need to impose $ \omega a \ll 1, \, \omega d \ll 1$, and $  \omega k \ll 1$. This also implies that $\omega q \ll 1$ and $  \omega p \ll 1$. Within this approximation, the radial equation (\ref{radial1}) reduces to 
\begin{equation}
	\left[ \partial_r \left(\Delta \partial_r\right) + \frac{r_+ - r_-}{r - r_+} A + \frac{r_+ - r_-}{r - r_-} B + C \right]  R(r) = 0 , \label{radeq} \
\end{equation} 
where
\begin{eqnarray}
&&A=\frac{\left[(r_+^2-d^2-k^2+a^2)\omega - am \right]^2}{(r_+ -r_-)^2},\nonumber\\
&&B=-\frac{\left[(r_-^2-d^2-k^2+a^2)\omega - am \right]^2}{(r_+ -r_-)^2}, \nonumber\\
&&C= -K_h, \label{eq:ABC}\
\end{eqnarray}
where we will take $K_h= h(h+1)$. This form of radial wave equation is very fruitful to reveal the hidden conformal symmetry. It is important to note also that we will not explore the angular wave equation because this does not have $ SL(2,R)\times SL(2,R) $ isometry, yet $ SU(2)\times SU(2) $ isometry \cite{LoweSkanataPRD2014}. So, in this case, we do not need to go further exploring the angular equation. This angular equation, somehow, is similar to those of Kerr black hole \cite{GuicaPRD2009}. The $ SL(2,R)\times SL(2,R) $ isometry is crucial to apply Cardy's growth of states from 2D CFT.

Secondly, after finding the radial equation in the low-frequency approximation, to reveal the hidden conformal symmetry, it is worth to perform the following coordinate transformations for the generic non-extremal black holes 
\begin{eqnarray}
&&\omega^+ = \sqrt{\frac{r-r_+}{r-r_-}}\mathrm{e}^{2\pi T_R \phi + 2 n_R t}, \nonumber\label{eq:con1}\\
&&\omega^- = \sqrt{\frac{r-r_+}{r-r_-}}\mathrm{e}^{2\pi T_L \phi + 2 n_L t}, \nonumber\label{eq:con2}\\
&&y = \sqrt{\frac{r_+ -r_-}{r-r_-}}\mathrm{e}^{\pi (T_L +T_R) \phi + (n_L + n_R) t}. \label{eq:con3}\
\end{eqnarray}
We can define three locally conformal operators in terms of the new conformal coordinates $\omega^+,\,\omega^-$ and $y$ as
\begin{eqnarray}
	&& H_1 = i \partial_+, \label{vec1}\\
	&& H_0 = i \left(\omega^{+}\partial_+ + \frac{1}{2}y\partial_y \right), \label{vec2}\\
	&& H_{-1} = i \left(\omega^{+2}\partial_+ + \omega^{+}y\partial_y - y^2 \partial_- \right), \label{vec3}
\end{eqnarray}
as well as 
\begin{eqnarray}
	&& \bar{H}_1 = i \partial_-, \label{vvec1}\\
	&& \bar{H}_0 = i\left(\omega^{-}\partial_- + \frac{1}{2}y\partial_y \right), \label{vvec2}\\
	&& \bar{H}_{-1} = i \left(\omega^{-2}\partial_- + \omega^{-}y\partial_y - y^2 \partial_+ \right). \label{vvec3}
\end{eqnarray}
Note that $\partial_\pm = \frac{\partial}{\partial \omega^\pm}$. A set of operators given in Eqs. (\ref{vec1})-(\ref{vec3}) satisfies the $ SL(2,R) $ Lie algebra as the follows
\begin{eqnarray}
	\left[H_0,H_{\pm 1} \right] = \mp iH_{\pm 1}, ~~~ \left[H_{-1},H_1 \right]=-2iH_0,
\end{eqnarray}
while other set of operators (\ref{vvec1})-(\ref{vvec3}) also forms $ SL(2,R) $ algebra. From any of two sets of operators, the quadratic Casimir operator can be formed as
\begin{eqnarray}
	\mathcal{H}^2 &=& \bar{\mathcal{H}}^2 = - H_0^2 + \frac{1}{2}(H_1 H_{-1} + H_{-1} H_{1})\nonumber\\
	 &=& \frac{1}{4}(y^2 \partial_y^2 - y\partial_y)+y^2\partial_+ \partial_-. \label{eq:quadraticCasimir}\
\end{eqnarray}
It is easier to notice the relation between the quadratic Casimir operator (\ref{eq:quadraticCasimir}) and the radial equation (\ref{radeq}) by bringing it back in terms of coordinates $ (t,r,\phi) $. In old coordinates, the quadratic Casimir operator  is given by
\begin{eqnarray}
	\mathcal{H}^2 &=& (r-r_+)(r-r_-) \partial_r^2 + (2r -r_+ -r_-)\partial_r \nonumber\\
	&+& \frac{r_+ -r_-}{r-r_-}\left(\frac{n_L-n_R}{4\pi G}\partial_\phi -\frac{T_L-T_R}{4G}\partial_t \right)^2 \nonumber\\
	&-& \frac{r_+ -r_-}{r-r_+}\left(\frac{n_L+n_R}{4\pi G}\partial_\phi -\frac{T_L+T_R}{4G}\partial_t \right)^2 , \label{CAS}
\end{eqnarray}
where $ G = n_L T_R - n_R T_L $.

We obtain the hidden conformal symmetry on the radial equation by comparing the radial equation (\ref{radeq}) and the Casimir operator (\ref{CAS}) that represents $SL(2,R)\times SL(2,R)$ isometry. We find that the radial equation (\ref{radeq}) can be rewritten in terms of the quadratic Casimir operator as given by
\begin{eqnarray}
	\mathcal{H}^2 R(r)=\bar{\mathcal{H}}^2 R(r)= -C R(r). \label{quadraticCasimirEq}
\end{eqnarray}
This radial equation stores the information of the temperatures and charges of 2D CFT given as follows
\begin{equation}
	n_L = -\frac{1}{2(r_+ + r_-)},~~~ n_R =0, \label{eq:nCFT}
\end{equation}
\begin{equation}
	T_L = \frac{r_+^2 + r_-^2+2(a^2-d^2-k^2)}{4\pi a(r_+ + r_-)},~~~ T_R =\frac{r_+ - r_-}{4\pi a}. \label{eq:tempCFT}
\end{equation}
For the extremal black hole, the non-vanishing temperature is only $T_L$ as given in Eq. (\ref{eq:TLKS}).
The temperature identification indicates the broken symmetry by the periodic identification $\phi \sim \phi +2\pi $ on the vector fields (\ref{vec1})-(\ref{vvec3}). This represents the spontaneously broken $SL(2,R)\times SL(2,R)$ symmetry into $U(1)\times U(1)$ subgroup.

To compute the entropy from CFT, we assume that the central charge of the extremal black holes connects smoothly with the non-extremal one \cite{SaktidyonicKerrSen} like in the original paper of hidden conformal symmetry \cite{Castro2010}. However, for future study, one can extend the calculation of the central charge using soft-hair method as given in Ref. \cite{HacoJHEP2018} where the result shall match with that of the extremal one given in Ref.~\cite{SaktidyonicKerrSen}. So, from  the central charge (\ref{eq:cLgdyonicKS}) we can rewrite for non-extremal one as follows
\begin{equation}
	c_L=c_R=12am_\pm= 6a(r_+ +r_-).\label{eq:cLcRdKS}
\end{equation}
Inserting this central charges together with the temperatures (\ref{eq:tempCFT}) to Cardy entropy formula
\begin{equation}
	S_{CFT} = \frac{\pi^2}{3}(c_L T_L + c_R T_R), \label{eq:cardyentropy}\
\end{equation}
one can find the entropy of non-extremal dyonic Kerr-Sen black hole. This entropy in given by
\begin{equation}
	S_{CFT}=\pi(r_+^2 -d^2 -k^2 +a^2). \label{eq:SCFTKS}
\end{equation}
This entropy will also exactly match with the entropy of non-extremal Kerr-Sen black hole when $p=0$ \cite{WuWuWuYuPRD2021} and matches with non-extremal Kerr black hole when $q=p=0$ \cite{Castro2010}. This further supports the results from extremal black holes stating that non-extremal dyonic Kerr-Sen black hole is holographically dual with 2D CFT.

\subsection{Gauged Dyonic Kerr-Sen black hole}\label{hiddengauged}
After showing the hidden conformal symmetry on the dyonic Kerr-Sen black, it is fascinating also to extend this computation for its gauged family or when the cosmological constant is present. We just need to employ the similar scalar wave equation (\ref{KG1}) and its ansatz (\ref{phi-expand1}) with background metric (\ref{eq:dKSAdSmetricshift}). Plugging Eq. (\ref{phi-expand1}) into Eq. (\ref{KG1}), leads to two differential equations i.e., the angular $S(\theta)$ and radial $R(r)$ wave functions,
\begin{eqnarray}
&& \left[\frac{1}{\sin\theta} \partial_\theta (\sin\theta \partial_\theta )-  \frac{m^2\Xi^2}{\sin^2\theta} +\frac{2am\omega\Xi- a^2\omega^2 \sin^2\theta}{\Delta_\theta} \right] S(\theta) \nonumber\\
&&=- K_h S(\theta)  \label{angulargauged1}, \
\end{eqnarray}
\begin{equation} 
	\bigg[ \partial_r (\Delta \partial_r) + \frac{\left[ (r^2 -d^2 -k^2 +a^2) \omega - am\Xi \right]^2}{\Delta} - K_{h} \bigg] R(r) = 0\label{radialgauged1}, \
\end{equation}
which are different from non-gauged case including separation constant $K_{h}$.

Similarly, the hidden conformal symmetries can be seen when we look at the near region and consider the low-frequency limit. These assumptions allow us to simplify the radial equation. Nevertheless, we need an extra treatment for this gauged family because the function $\Delta$ is quartic because there exist cosmological horizons beside the inner and outer horizons. It is compulsory to approximate $\Delta$ in the near-horizon region. In the near-horizon region, we can obtain the following approximation
\begin{equation}
	\Delta \simeq \upsilon (r-r_+)(r-r_*), \label{eq:aprroxdelta}\
\end{equation}
where $\upsilon$ is given in (\ref{eq:upsilonKSAdS}) and
\begin{eqnarray}
	r_* &=& r_+ - \frac{1}{\upsilon r_+}\bigg[\frac{2r_+^2(2r_+^2 -2d^2-2k^2 +a^2+l^2)}{l^2} \nonumber\\
	&-&\frac{(r_+^2 -d^2-k^2 +a^2)(r_+^2 -d^2-k^2 +l^2)}{l^2}\nonumber\\
	&+& q^2+p^2\bigg].\
\end{eqnarray}
Note that $r_*$ is not the inner horizon. However, when the gauge coupling constant satisfies $1/l^2 \sim 0$, this is exactly similar.
Now the radial equation (\ref{radialgauged1}) is simplified as given by
\begin{equation}
	\left[ \partial_r \left(\frac{\Delta}{\upsilon} \partial_r\right) + \frac{r_+ - r_*}{r - r_+} A_s + \frac{r_+ - r_*}{r - r_*} B_s + C_s \right]  R(r) = 0 , \label{radeqgauged} \
\end{equation} 
where
\begin{eqnarray}
&&A_s=\frac{\left[(r_+^2-d^2-k^2+a^2)\omega - am\Xi \right]^2}{\upsilon^2(r_+ -r_*)^2},\nonumber\\
&&B_s=-\frac{\left[(r_*^2-d^2-k^2+a^2)\omega - am\Xi \right]^2}{\upsilon^2(r_+ -r_*)^2}, \nonumber\\
&&C_s= -\frac{K_h}{\upsilon}\label{eq:ABCgauged}\
\end{eqnarray}
It is worth to note that this radial equation can be used to calculate the entropy. However, in order to compute the scattering, we need to consider also the asymptotic region which in this case, we cannot use this radial equation.

After finding the radial equation in the low-frequency approximation,  we will perform again the coordinate transformations. These coordinate transformations are not so different with the non-gauged family since we just need to use $r_*$ instead of using $r_-$. For the generic non-extremal black holes, we have the following coordinate transformations
\begin{eqnarray}
&&\omega^+ = \sqrt{\frac{r-r_+}{r-r_*}}\mathrm{e}^{2\pi T_R \phi + 2 n_R t}, \nonumber\label{eq:con1gauged}\\
&&\omega^- = \sqrt{\frac{r-r_+}{r-r_*}}\mathrm{e}^{2\pi T_L \phi + 2 n_L t}, \nonumber\label{eq:con2gauged}\\
&&	y = \sqrt{\frac{r_+ -r_*}{r-r_*}}\mathrm{e}^{\pi (T_L +T_R) \phi + (n_L + n_R) t}. \label{eq:con3gauged}\
\end{eqnarray}
First set of locally conformal operators in terms of the new conformal coordinates is similar as given by  (\ref{vec1})-(\ref{vec3}) that satisfies the $ SL(2,R) $ Lie algebra. The other set is also similar as (\ref{vvec1})-(\ref{vvec3}) that also forms $ SL(2,R) $ algebra. From any of two sets of operators, again, we can form a quadratic Casimir operator (\ref{eq:quadraticCasimir}).

It is obvious that we have revealed the hidden conformal symmetry on the radial equation by comparing the radial equation (\ref{radeqgauged}) and the Casimir operator (\ref{CAS}) that represents $SL(2,R)\times SL(2,R)$ isometry for this gauged family. The remaining differences from non-gauged case are the stored information which are the temperatures and charges of 2D CFT. The temperatures and charges of 2D CFT for the gauged dyonic Kerr-Sen solution are given by the following
\begin{equation}
	n_L = -\frac{\upsilon}{2(r_+ + r_*)},~~~ n_R =0, \label{eq:nCFTgauged}
\end{equation}
\begin{equation}
	T_L = \frac{\upsilon[r_+^2 + r_*^2+2(a^2-d^2-k^2)]}{4\pi a(r_+ + r_*)\Xi},~~~ T_R =\frac{\upsilon(r_+ - r_*)}{4\pi a\Xi}. \label{eq:tempCFTgauged}
\end{equation}
For the extremal black hole, the non-vanishing temperature is only $T_L$ as given in Eq. (\ref{eq:TLgeneral}). Similarly with non-gauged family, the temperature identification indicates the broken symmetry spontaneously from $SL(2,R) \times SL(2,R)$ into $U(1)\times U(1)$ subgroup by the periodic identification of $\phi$.

Likewise the non-gauged case, the central charges are assumed to be connected smoothly with that of extremal case which is given in Eq. (\ref{eq:cLgeneral}) that is recalculated as given in Eq. (\ref{eq:cLtot}). Hence, we can find that
\begin{equation}
	c_L=c_R= \frac{6a(r_+ +r_*)}{\upsilon}.\label{eq:cLcRgauged}
\end{equation}
For non-gauged case, we know that $\upsilon = 1, r_*=r_-$. So, in this case, the central charges reduce to those of dyonic Kerr-Sen black hole. By employing this central charges together with the temperatures (\ref{eq:tempCFTgauged}) to Cardy entropy formula (\ref{eq:cardyentropy}), one can obtain the entropy of non-extremal gauged dyonic Kerr-Sen black hole as given by
\begin{equation}
	S_{CFT}=\frac{\pi}{\Xi}(r_+^2 -d^2 -k^2 +a^2). \label{eq:SCFTgauged}
\end{equation}
We can also recover the entropy of non-extremal gauged Kerr-Sen black hole when $p=0$ \cite{WuWuWuYuPRD2021} and the non-extremal Kerr-AdS black hole when $q=p=0$. Again, this result supports the conjecture that non-extremal gauged dyonic Kerr-Sen black hole is holographically dual with 2D CFT.

\section{Scattering cross-section on non-extremal background}\label{sec:scat}
Beside the entropy calculation, another probe to prove the Kerr/CFT correspondence conjecture is to calculate the absorption cross-section. The calculation of the absorption cross-section from 2D CFT shall match the calculation from gravity. In this section, we will prove that matching on  the absorption cross-section.

\subsection{Scattering off dyonic Kerr-Sen black hole}

In this subsection, we consider the absorption cross-section of scalar probes in the background of the generic non-extremal dyonic Kerr-Sen black hole. So, we need to find the solution to the scalar wave equation (\ref{radeq}). In order to solve that equation, we require to use the following radial coordinate transformation
\begin{equation}
	z=\frac{r-r_+}{r-r_-}. \label{eq:ztor}
\end{equation}
This implies that when $r_+\leq r \leq \infty$, we have $0 \leq z \leq 1$. In this new coordinate, it is easy to find that $1-z=(r_+ -r_-)/(r-r_-)$. Now the radial part of the wave equation becomes 
\begin{equation}
	\left[	z(1-z)\partial^2_z + (1-z)\partial_z +\frac{A}{z}+B +\frac{C}{1-z} \right] R(z)=0, \label{radeqz}
\end{equation}
where the constants $A,B, C$ are given by (\ref{eq:ABC}). Above equation has the ingoing and outgoing solutions and can be solved by hypergeometric function as given by
\begin{eqnarray}
	&& R^{in}(z)=z^{-i\sqrt{A}}(1-z)^{(1+l)} ~_2F_1(a_s,b_s;c_s;z),\label{eq:ingoingsol}\\
	&& R^{out}(z)=z^{i\sqrt{A}}(1-z)^{(1+l)} ~_2F_1(a_s^*,b_s^*;c_s^*;z),\label{eq:outgoingsol}\
\end{eqnarray}
respectively, where $a_s=1+h -i(\sqrt{A}+\sqrt{-B})$, $b_s=1+h -i(\sqrt{A}-\sqrt{-B})$, and $c_s = 1-2i\sqrt{A}$. To study the absorption cross-section for a wave coming from infinity towards the outer event horizon, the ingoing solution will be considered. We want to see the asymptotic behavior of the solution at the matching region $r\gg M$ but still satisfying near region ($r\ll 1/\omega$). In order to do so, we can use the following relation
\begin{eqnarray}
&&~_2F_1(a_s,b_s;c_s;z)\nonumber\\
&=&\frac{\Gamma(c_s)\Gamma(c_s-a_s-b_s)}{\Gamma(c_s-a_s)\Gamma(c_s-b_s)} ~_2F_1(a_s,b_s;a_s+b_s-c_s;1-z)\nonumber\\
&+&(1-z)^{c_s-a_s-b_s} \frac{\Gamma(c_s)\Gamma(a_s+b_s-c_s)}{\Gamma(a_s)\Gamma(b_s)}\nonumber\\
&\times&~_2F_1(c_s-a_s,c_s-b_s;c_s-a_s-b_s;1-z),
\label{eq:gammarel}\
\end{eqnarray}
and $~_2F_1(a_s,b_s;c_s;0)=1$. Hence, in asymptotic region $z\rightarrow 1$, we have
\begin{equation}
R^{in}(r\gg M) \sim D_0 r^h +D_1 r^{-1-h}, \label{eq:asymptotsol}\
\end{equation}
where 
\begin{equation}
	D_0 = \frac{\Gamma(c_s)\Gamma(1+2h)}{\Gamma(a_s)\Gamma(b_s)}, ~~~ D_1 = \frac{\Gamma(c_s)\Gamma(-1-2h)}{\Gamma(c_s-a_s)\Gamma(c_s -b_s)}.\label{eq:gammadefinition}\
\end{equation}
The essential part of the absorption cross-section can be read out directly from the coefficient $D_0$, namely,
\begin{equation}
	P_{abs} \sim \left| D_0 \right|^{-2} \sim \sinh \left( {2\pi A^{1/2} } \right){\left| {\Gamma \left(a_s  \right)}  {\Gamma \left(b_s \right)} \right|^2 }\label{eq:Pabs}.
\end{equation}
Note that the constant $D_1$ is suppressed by the constant $D_0$, so we can ignore $D_1$ in the absorption cross-section above. For more detail explanation, we refer to see Ref. \cite{BredbergStroJHEP2010}.

After deriving the absorption cross-section from gravity calculation, it is time to derive it from CFT side. The absorption cross-section from CFT is given by \cite{BredbergStroJHEP2010}
\begin{eqnarray}
	P_{abs} &\sim & {T _L}^{2h_L - 1} {T _R}^{2h_R - 1} \sinh \left( {\frac{{{\tilde{\omega}} _L }}{{2{T _L} }} + \frac{{{\tilde{\omega}} _R }}{{2{T _R} }}} \right)\nonumber\\
&\times& \left| {\Gamma \left( {h_L + i\frac{{{\tilde{\omega}} _L }}{{2\pi {T _L} }}} \right)} {\Gamma \left( {h_R + i\frac{{{\tilde{\omega}} _R }}{{2\pi {T _R} }}} \right)} \right|^2.\label{eq:PabsCFT}\
\end{eqnarray}
We already have the left- and right- moving temperatures of the dyonic Kerr-Sen black hole given in Eq. (\ref{eq:tempCFT}). The conformal weights of the operator dual to the scalar
field are given by
\begin{equation}
	h_L=h_R = h+1. \label{eq:confweight}\
\end{equation}
The last, to find agreement between (\ref{eq:Pabs}) and (\ref{eq:PabsCFT}), we need to identify the proper left and right frequencies $\omega _L,\omega _R$. In order to do so, we consider the first law of thermodynamics for the general charged rotating black holes, namely,
\begin{equation} \label{eq:BHthermoLaw}
	T_H \delta S_{BH} = \delta M - \Omega _H \delta J - \Phi _H \delta Q-\Psi _H \delta P,
\end{equation} 
where $T_H$, $S_{BH}$, $\Omega_H$, $\Phi_H$, and $\Psi_H$ are given by Eqs.  (\ref{eq:THKS})-(\ref{eq:potpsiKS}). Nevertheless, it is crucial that for neutral scalar probe we use in this computation, we have $\delta Q= \delta P=0$.
Furthermore, we can vary the Cardy formula (\ref{eq:cardyentropy}) in terms of conjugate charges $E_L,E_R$ as given by
\begin{equation} \label{eq:delSCFT}
	\delta S_{CFT} = \frac{{\delta E_L }}{{T_L }} + \frac{{\delta E_R }}{{T_R }}.
\end{equation}
One can find the conjugate charges via $\delta S_{BH}=\delta S_{CFT}$. Note that we do not consider the correction of the entropy, so we can apply that relation \cite{Compere2017}. Using the identification $\omega= \delta M$ and $m = \delta J$ yields to the identification of $\delta E _{L,R}$ as $\omega_{L,R}$. By equating the 
variations of entropy in (\ref{eq:BHthermoLaw}) and (\ref{eq:delSCFT}), we obtain that
\begin{eqnarray}
&&\delta E _{L}=\omega _{L} = \frac{r_+^2 + r_-^2 +2(a^2-d^2-k^2)}{2a}\omega,\nonumber\\
&&\delta E _{R}=\omega _{R} = \omega _{L}-m.  \label{eq:omegaCFT} \
\end{eqnarray}
These identifications are needed in order to match the Bekenstein-Hawking entropy and Cardy entropy formula. Furthermore, one can check that these identifications are consistent with the result for the extremal case. Using this result, we finally can write
\begin{equation}
	a_s =h_R-i\frac{\omega_R}{2\pi T_R}, ~ b_s =h_L-i\frac{\omega_L}{2\pi T_L}, ~ 2\pi A^{1/2}=\frac{\omega_L}{2T_L}+\frac{\omega_R}{2T_R}. 
\end{equation}
So, we find the final agreement between absorption cross-section between gravity calculation (\ref{eq:Pabs}) and CFT calculation (\ref{eq:PabsCFT}). Those are the finite temperature absorption cross-section of an operator dual to the neutral probe scalar field in 2D CFT to dyonic Kerr-Sen black hole.

\subsection{Scattering off gauged dyonic Kerr-Sen black hole}

Now we consider scalar probes in the background of the generic non-extremal gauged dyonic Kerr-Sen black hole. So, we need to find the solution of the scalar wave equation (\ref{radeqgauged}). In order to solve the equation, we use the following coordinate transformation
\begin{equation}
	z=\frac{r-r_+}{r-r_*}. \label{eq:ztorgauged}
\end{equation}
In this new coordinate, the radial wave equation becomes 
\begin{equation}
	\left[	z(1-z)\partial^2_z + (1-z)\partial_z +\frac{A_s}{z}+ B_s +\frac{C_s}{1-z} \right] R(z)=0, \label{radeqzgauged}
\end{equation}
where the constants $A_s, B_s, C_s$ are given by (\ref{eq:ABCgauged}). Similarly, that radial wave equation can be solved by hypergeometric function where the ingoing solution is also given by Eq. (\ref{eq:ingoingsol}) where now $a_s=1+h -i(\sqrt{A_s}+\sqrt{-B_s})$, $b_s=1+h -i(\sqrt{A_s}-\sqrt{-B_s})$, and $c_s = 1-2i\sqrt{A_s}$. It is important to note that the constant $h$ is different with non-gauged case where it should satisfy
\begin{equation}
	h=\frac{1}{2}\left(-1+ \sqrt{1-4C_s} \right).\label{eq:hgauged}
\end{equation}

Note that the radial equation (\ref{radeqzgauged}) holds only at the very near-horizon region. Although we can find the scalar radial solution, we cannot study the asymptotic behavior of the radial wave function because the expansion of $\Delta$ to the quadratic order in near-horizon region is problematic. Nonetheless, if we consider the near-horizon region of the near-extremal black
holes, the same treatment can still be employed. From the definition of $z$ on (\ref{eq:ztorgauged}), only when $r_+ \simeq r_*$, we do not need to move very far from the horizon to find $z \rightarrow 1$. For this condition, the discussion of the scattering amplitudes for the near-extremal gauged dyonic Kerr-Sen black hole will be given in the next subsection. However, we can still study conjugate charges $E_L,E_R$ for this generic non-extremal gauged Kerr-Sen black hole.

In order to compute the conjugate charges, we consider the first law of thermodynamics for the general gauged charged rotating black holes, namely,
\begin{equation} \label{eq:BHthermoLawgauged}
	T_H \delta S_{BH} = \delta M - \Omega _H \delta J - \Phi _H \delta Q-\Psi _H \delta P -V\delta \mathcal{P},
\end{equation} 
where above quantities are given by Eqs. (\ref{eq:MJQPKSAdS}) - (\ref{eq:VdKSAdS}). Note that for neutral scalar probe, we have $\delta Q= \delta P=\delta\mathcal{P}=0$. Moreover, the variation Cardy formula in terms of conjugate charges $E_L,E_R$ is given by (\ref{eq:delSCFT}). Again, one can find the conjugate charges via $\delta S_{BH}=\delta S_{CFT}$. Using the identification $\delta M$ as $\omega$ and $\delta J$ as $m$ yields to the  identification of $\delta E _{L,R}$ as $\omega_{L,R}$. By equating the 
variations of entropy in (\ref{eq:delSCFT}) and (\ref{eq:BHthermoLaw}), we obtain that
\begin{eqnarray}
&&\delta E _{L}=\omega _{L} = \frac{r_+^2 + r_*^2 +2(a^2-d^2-k^2)}{2a\Xi}\omega, \nonumber\\
&&\delta E _{R}=\omega _{R} = \omega _{L}-m.  \label{eq:omegaCFTgauged} \
\end{eqnarray}
So, we find the left and right frequencies of 2D CFT for generic non-extremal gauged dyonic Kerr-Sen black holes.

\subsection{Superradiant scattering off gauged dyonic Kerr-Sen black hole}
As we have mentioned in the previous subsection, we cannot study the scattering issue on generic non-extremal gauged dyonic Kerr-Sen black hole due to the breaking down of near-horizon approximation on asymptotic region. However, this might be avoided by working on near-extremal case.
For the near-horizon region, where we can expand the metric function $ \Delta $  by the quadratic order in $ (r-r_+) $ as (\ref{eq:aprroxdelta}), we use the radial wave equation (\ref{radeqgauged}) to find the scattering cross-section.  So, for a near-extremal black hole, it is allowed to consider the following near-extremal coordinate transformations from black hole coordinates $(r,t,\phi)$ to $(y,\tau,\varphi)$, as
\begin{equation}
r = \frac{r_+ + r_*}{2}+\lambda r_0 y, ~~r_+ -r_* = \mu \lambda r_0,\nonumber\
\end{equation}
\begin{equation}
t = \frac{r_0 \Xi}{\lambda}\tau, ~~ \phi= \varphi + \frac{\Omega_H r_0 \Xi}{\lambda}\tau, \label{nearextremaltransformation} \
\end{equation}
where $r_0^2=r_+^2 + a^2 -d^2 -k^2$ and $ \lambda \rightarrow 0 $ shows the near-horizon limit and $\mu$ is the near-extremality parameter. We also need to consider the scalar probe with frequencies around the superradiant bound $ \omega_s = m\Omega_H $, given by
\begin{equation}
	\omega = \omega_s +\hat{\omega}\frac{\lambda}{r_0}.
\end{equation}
The limit $\lambda \rightarrow 0$ implies the superradiant bound. We can re-write the radial equation (\ref{radeqgauged}) by
\begin{equation}
	\left[\partial_y\left(y-\frac{\mu}{2}\right)\left(y+\frac{\mu}{2}\right)\partial_y +\frac{A_t}{y-\frac{\mu}{2}}+\frac{B_t}{y+\frac{\mu}{2}} + C_t\right]R(y) =0,\label{radialequationnearext}
\end{equation}
where
\begin{equation}
	A_t = \frac{\hat{\omega}^2}{\upsilon^2\mu}, ~~ B_t = -\frac{\mu}{\upsilon^2} \left(\frac{\hat{\omega}}{\mu}-2 m\Omega_H r_+ \right)^2, ~~C_t=C_t(\hat{\omega}),\
\end{equation}
and $ C_t $ is the separation constant which is distinct with the one when we have not deemed the superradiant bound. Then we can perform the change of coordinate from $y$ to $z$, by 
$ z =\frac{y-\mu/2}{y+\mu/2} $, where the radial equation (\ref{radialequationnearext}) becomes
\begin{equation}
	\left[z(1-z)\partial_z^2 + (1-z)\partial_z +\frac{\hat{A_t}}{z}+\hat{B_t}+\frac{C_t}{1-z}\right]R(z) =0,\label{radialequationnearext1}
\end{equation}
where
\begin{eqnarray}
	\hat{A_t} = \frac{\hat{\omega}^2}{\upsilon^2\mu ^2}, ~~~\hat{B_t} = -\frac{1}{\upsilon^2} \left(\frac{\hat{\omega}}{\mu}-2m\Omega_H r_+\right)^2. 
\end{eqnarray}

Above form of radial equation is similar to that of non-gauged case (\ref{radeqz}).
The ingoing solution to differential equation (\ref{radialequationnearext1}) is also given by hypergeometric function
\begin{equation}
	R(z)= z^{-i\sqrt{\hat{A_t}}}(1-z)^{1+h}~_2F_1(a_s,b_s;c_s;z), \label{eq:ingoingsolgauged}\
\end{equation}
with the parameters 
$a_s=1+h -i(\sqrt{\hat{A}_t}+\sqrt{-\hat{B}_t})$, $b_s=1+h -i(\sqrt{\hat{A}_t}-\sqrt{-\hat{B}_t})$, and $c_s = 1-2i\sqrt{\hat{A}_t}$. For this superradiant case, the relation between $h$ and $C_t$ is given by
\begin{equation}
	h=\frac{1}{2}\left(-1+ \sqrt{1-4C_t} \right).\label{eq:h}
\end{equation}
For the asymptotic region of the radial coordinate $r$ (or equivalently $y\gg\mu/2 $), where  $z \sim 1$, the solution (\ref{eq:ingoingsolgauged}) reduces to
\begin{eqnarray}
	R(y) \sim D_0 y^{h}+D_1 y^{-1-h}, \label{eq:asymptotsol0}\
\end{eqnarray}
where
\begin{equation}
	D_0 = \frac{\Gamma(c_s)\Gamma(1+2h)}{\Gamma(a_s)\Gamma(b_s)}, ~~~ D_1 = \frac{\Gamma(c_s)\Gamma(-1-2h)}{\Gamma(c_s-a_s)\Gamma(c_s -b_s)}.\label{eq:gammadefinition0}\
\end{equation}
Note that $h+1$ is the conformal weight of the scalar field. 

For the coefficient (\ref{eq:gammadefinition0}), we find the absorption cross-section of the scalar fields as 
\begin{equation}
	P_{abs} \sim \left| D_0 \right|^{-2} \sim \sinh \left( {2\pi \hat{A}_t^{1/2} } \right){\left| {\Gamma \left(a_s  \right)}  {\Gamma \left(b_s \right)} \right|^2 }\label{eq:Pabsgauged}.
\end{equation}
To further support the correspondence between the near-extremal gauged Kerr-Sen black hole and 2D CFT, we show that the absorption cross-section for the scalar fields  (\ref{eq:Pabsgauged}) can be obtained from the absorption cross-section in a 2D CFT (\ref{eq:PabsCFT}).
To find the agreement between (\ref{eq:PabsCFT}) and (\ref{eq:Pabsgauged}), we require to choose proper left and right frequencies $\omega_L,\omega_R$. In order to do so, we can consider the first law of thermodynamics given in Eq. (\ref{eq:BHthermoLawgauged}) and compare with Eq. (\ref{eq:delSCFT}). For near-extremal and near-horizon gauged dyonic Kerr-Sen family, we obtain
\begin{equation}
	\omega_L = m, ~~~ \omega_R = \frac{r_0}{a\Xi}\left(\hat{\omega}-\mu m\Omega_H r_+ \right),\label{eq:omegasuper}\
\end{equation}
while the temperatures and conformal weights are now given by
\begin{equation}
	T_L = \frac{\upsilon}{4\pi\Omega_H r_+}, ~~~ T_R = \frac{\upsilon r_0}{4\pi a\Xi}\lambda\mu, ~~~ h_L=h_R =1+h.\label{eq:tempsuper}\
\end{equation}

This result is in a good match with the CFT prediction (\ref{eq:PabsCFT}). This remarkable identifications (\ref{eq:omegasuper})-(\ref{eq:tempsuper}) are exactly the same as the
ones (\ref{eq:tempCFTgauged}) and (\ref{eq:omegaCFTgauged}), which were obtained respectively from the conformal coordinate transformation in the low-frequency limit and the first law of thermodynamics for generic non-extremal gauged dyonic Kerr-Sen black hole. This gives rise to another nontrivial evidence to support the Kerr/CFT correspondence for the black holes in EMDA theory.

\subsection{Real-time correlator}
More on supporting Kerr/CFT correspondence, one can also compute the real-time correlator. The asymptotic behaviors of the scalar field with ingoing boundary condition on the background of dyonic Kerr-Sen black hole (\ref{eq:asymptotsol}) and on gauged dyonic Kerr-Sen black hole (\ref{eq:asymptotsol0}) indicate that two coefficients possess different roles. $D_0$ indicates the source while $D_1$ indicates the response. Hence, the two-point retarded correlator is simply \cite{ChenChuCorrelatorJHEP2010,ChenLongCorrelatorJHEP2010}
\begin{equation}
	G_R \sim \frac{D_1}{D_0} = \frac{\Gamma(-1-2h)}{\Gamma(1+2h)}\frac{\Gamma(a_s)\Gamma(b_s)}{\Gamma(c_s-a_s)\Gamma(c_s-b_s)},\label{eq:retardedcorr}\
\end{equation}
on dyonic Kerr-Sen black hole's background. For gauged family, the two-point retarded correlator has similar form as (\ref{eq:retardedcorr}), however, the values of $h,a_s,b_s,c_s$ are different. From Eq. (\ref{eq:retardedcorr}), it is easy to check that
\begin{equation}
	G_R \sim \frac{\Gamma(h_L - i\frac{\omega_L}{2T_L})\Gamma(h_R - i\frac{\omega_R}{2T_R})}{\Gamma(1-h_L - i\frac{\omega_L}{2T_L})\Gamma(1-h_R - i\frac{\omega_R}{2T_R})}.
\end{equation}
Then we can write above two-point retarded correlator into
\begin{eqnarray}
	G_R &\sim& \sin\left(\pi h_L+i\frac{\omega_L}{2T_L}\right)\sin\left(\pi h_R+i\frac{\omega_R}{2T_R}\right)\nonumber\\
	&\times&\bigg|\Gamma\left(h_L + i\frac{\omega_L}{2T_L}\right)\Gamma\left(h_R + i\frac{\omega_R}{2T_R}\right)\bigg|^2, \label{eq:correlator}
\end{eqnarray}
by using the relation $\Gamma(z)\Gamma(1-z)=\pi/\sin(\pi z)$. Since $h_L,h_R$ are integers, so we have
\begin{eqnarray}
&&\sin\left(\pi h_L+i\frac{\omega_L}{2T_L}\right)\sin\left(\pi h_R+i\frac{\omega_R}{2T_R}\right)\nonumber\\
&&= (-1)^{h_L+h_R}\sin\left(i\frac{\omega_L}{2T_L}\right)\sin\left(i\frac{\omega_R}{2T_R}\right). \label{eq:sinsorrelator}\
\end{eqnarray}

In CFT, the Euclidean correlator is given by
\begin{eqnarray}
G_E (\omega_{EL},\omega_{ER}) &\sim& T_L^{2h_L-1} T_R^{2h_R-1}e^{i\omega_{EL}/2T_L}e^{i\omega_{ER}/2T_R}\nonumber\\
&\times&\bigg|\Gamma\left(h_L + \frac{\omega_{EL}}{2T_L}\right)\Gamma\left(h_R + \frac{\omega_{EL}}{2T_R}\right)\bigg|^2.\nonumber\\
\
\end{eqnarray}
Where we can define the Euclidean frequencies $\omega_{EL}=i\omega_L$, and $\omega_{ER}=i\omega_R$. It is important noting that $G_E$ corresponds to the values of retarded correlator $G_R$. The retarded Green function $G_R$ is analytic on the upper half complex $\omega_{L,R}$-plane. The value of $G_R$ along the positive imaginary $\omega_{L,R}$-axis gives the following correlator
\begin{equation}
	G_E(\omega_{EL},\omega_{ER}) = G_R(i\omega_{L},i\omega_{R}), ~~~ \omega_{EL,ER}>0. \label{eq:correlatormatch}
\end{equation}
At finite temperature, $\omega_{EL,ER}$ should take discrete values of the Matsubara frequencies, given by
\begin{equation}
	\omega_{EL} =2\pi m_L T_L, ~~~ \omega_{ER}=2\pi m_R T_R, \label{eq:Matsubara}
\end{equation}
where $m_L$ and $m_R$ are half integers for fermionic modes and integers for bosonic modes. At these certain frequencies, the gravity computation for correlator (\ref{eq:correlator}) matches precisely with CFT result trough Eq. (\ref{eq:correlatormatch}) up to a numerical normalization factor.

\section{Conclusions}\label{sec:conc}
In this paper, we have explicitly extended the study of holography on the dyonic black holes in EMDA theory with vanishing and non-vanishing gauge coupling constant $l$. Specifically, we have constructed the hidden conformal invariance for the dyonic Kerr-Sen and gauged dyonic Kerr-Sen black holes. To reveal the hidden conformal invariance for both black holes, we have mainly deemed the near region and the low-frequency limit. Moreover, for gauged family, we also needed to approximate to the near-horizon region to obtain the expected radial equation. We have exhibited that both radial equations on different black hole's background possess $SL(2,R)\times SL(2,R)$ squared Casimir equation, denoting a local hidden conformal symmetry by the periodic identification $\phi \sim \phi +2\pi$. This finding suggests that both dyonic black holes are dual in 2D CFT with finite left- and right-moving temperatures. We have also proved the previous calculation of the central charge by considering the topological term and other correponding fields. By using this, we then obtained the matching between Cardy entropy formula and Bekenstein-Hawking entropy of both dyonic black holes. This denotes that generic non-extremal dyonic Kerr-Sen black hole and its gauged family are holographically dual to 2D CFT.

To further support the holography, the absorption cross-section of the scalar field perturbation has also been considered via gravity and CFT computations. In investigating the scattering issue, for gauged one, we could not use the near-horizon approximation for the generic non-extremal black hole. Nevertheless, we could apply that approximation when we consider the near extremality.
From our computation, the matching of the absorption cross-section has been exhibited to have a perfect agreement by identifying the arbitrary left and right CFT frequencies. It has been shown also the temperatures for near-extremal condition.

We have then performed the computation of retarded Green's function or the two-point correlator. We have found the perfect agreement between the retarded Green's functions and the CFT Euclidean correlators which are restricted by conformal invariance. It is worth noting also that the real-time correlator is closely related to the absorption cross-section where we could see from the imaginary part of the retarded Green's function at Matsubara frequencies.

\begin{acknowledgement}
\textbf{Acknowledgement} \\
We would like to thank the anonymous referee for valuable comments which help to improve our work considerably. This research is supported by the Second Century Fund (C2F), Chulalongkorn University, Thailand.\end{acknowledgement}

\appendix
\section{Effective Lagrangian of Low Energy Limit of Heterotic String Theory}
\label{app:1}
In this paper, we consider two different actions given by (\ref{eq:Lagrangian}) and its gauged one (\ref{eq:Lagrangian1}). The difference between those action is the presence of the potential term as shown in Eq. (\ref{eq:Lagrangian1}). In the case of $1/l^2=0$, the potential term containing cosmological constant will vanish. Interestingly, one can write the Lagrangian (\ref{eq:Lagrangian}) in another form, the effective Lagrangian of the low energy limit of the heterotic string theory \cite{WuWuWuYuPRD2021} given by
\begin{equation}
\mathcal{L}_{eff}=\sqrt{-g}\left(R-\frac{1}{2}(\partial\phi)^2-e^{-\phi}F^2-\frac{1}{12}e^{-2\phi}H^2\right), \label{eq:effLagrangian}
\end{equation}
where $H^2=H_{\mu\nu\rho}H^{\mu\nu\rho}$
is an antisymmetric tensor where it is defined by $H=d\mathcal{B}-A\wedge F/4=-e^{2\phi}\star d\chi$. For tis effective Lagrangian, we need to introduce a new field $\mathcal{B}$, the antisymmetric tensor field. The star operator represents Hodge duality. The term $A\wedge F/4$ is the Chern-Simons term. However, when the cosmological constant is present, one cannot write (\ref{eq:Lagrangian1}) into above action, so this is a special case in supergravity theory. Hence, the main difference between the effective action for dyonic Kerr-Sen solution and its gauged family is the potential term which contains cosmological constant that couples to the dilaton and axion fields.

%
%
%
%

\end{document}